\documentclass[conference,10pt]{IEEEtran}
\usepackage{amsmath, graphics,amssymb,epsfig,subfigure}

\usepackage{cite}
\usepackage{url}
\usepackage{color,soul}

\def\BibTeX{{\rm B\kern-.05em{\sc i\kern-.025em b}\kern-.08em
    T\kern-.1667em\lower.7ex\hbox{E}\kern-.125emX}}

\newcommand{\be}{\begin{equation}}

\newcommand{\ee}{\end{equation}}
\newcommand{\bea}{\begin{eqnarray}}
\newcommand{\eea}{\end{eqnarray}}
\newcommand{\bdp}{\begin{displaymath}}
\newcommand{\edp}{\end{displaymath}}
\usepackage{graphicx,amsmath}

\makeatletter
\def\hlinewd#1{%
\noalign{\ifnum0=`}\fi\hrule \@height #1 \futurelet \reserved@a\@xhline}
\newcommand{\hthickline}{\hlinewd{.8pt}}

\makeatother
\begin{document}

\title{\huge{Joint Design of Digital and Analog Processing for Downlink C-RAN with Large-Scale Antenna Arrays}}

\author{\IEEEauthorblockN{{$^1$Jaein Kim, $^2$Seok-Hwan Park, $^3$Osvaldo Simeone, $^1$Inkyu Lee and $^4$Shlomo Shamai (Shitz)}
\IEEEauthorblockA{$^1$School of Electrical Engineering, Korea University, Seoul, Korea\\$^2$Division of Electronic Engineering, Chonbuk National University, Jeonju, Korea\\$^3$Center for Wireless Information Processing, New Jersey Institute of Technology, Newark, New Jersey, USA\\$^4$Department of Electrical Engineering, Technion, Haifa, Israel\\Email: kji\_07@korea.ac.kr, seokhwan@jbnu.ac.kr, osvaldo.simeone@njit.edu, inkyu@korea.ac.kr, sshlomo@ee.technion.ac.il}}
}
\maketitle \thispagestyle{empty}
\begin{abstract}

In millimeter-wave communication systems with large-scale antenna arrays, conventional digital beamforming may not be cost-effective.
A promising solution is the implementation of hybrid beamforming techniques, which consist of low-dimensional digital beamforming followed by analog radio frequency (RF) beamforming.
This work studies the optimization of hybrid beamforming in the context of a cloud radio access network (C-RAN) architecture.
In a C-RAN system, digital baseband signal processing functionalities are migrated from remote radio heads (RRHs) to a baseband processing unit (BBU) in the "cloud" by means of finite-capacity fronthaul links.
Specifically, this work tackles the problem of jointly optimizing digital beamforming and fronthaul quantization strategies at the BBU, as well as RF beamforming at the RRHs, with the goal of maximizing the weighted downlink sum-rate.
Fronthaul capacity and per-RRH power constraints are enforced along with constant modulus constraints on the RF beamforming matrices.
An iterative algorithm is proposed that is based on successive convex approximation and on the relaxation of the constant modulus constraint. The effectiveness of the proposed scheme is validated by numerical simulation results.
\end{abstract}

\begin{IEEEkeywords}
Cloud-RAN, mmWave communication, hybrid beamforming.
\end{IEEEkeywords}

\section{Introduction} \label{sec:intro}

Millimeter-wave communication technology has the potential not only of alleviating the bandwidth shortage but also of enabling the deployment of large-scale antenna arrays (see, e.g., \cite{TSRa:13,ZPi:11,FSo:16}).
However, it may be impractical to equip every antenna of a large array with a radio frequency (RF) chain due to the cost considerations and hardware limitations.
Hybrid beamforming techniques, whereby the beamforming process consists of a low-dimensional digital bemforming followed by analog RF beamforming, has emerged as an effective means to address this problem (see, e.g., \cite{LLi:14,OEAy:14,AAl:15,XYu:16,FSo:16,CSL:16}).

\begin{figure}
\begin{center}
\includegraphics[width=3.5 in]{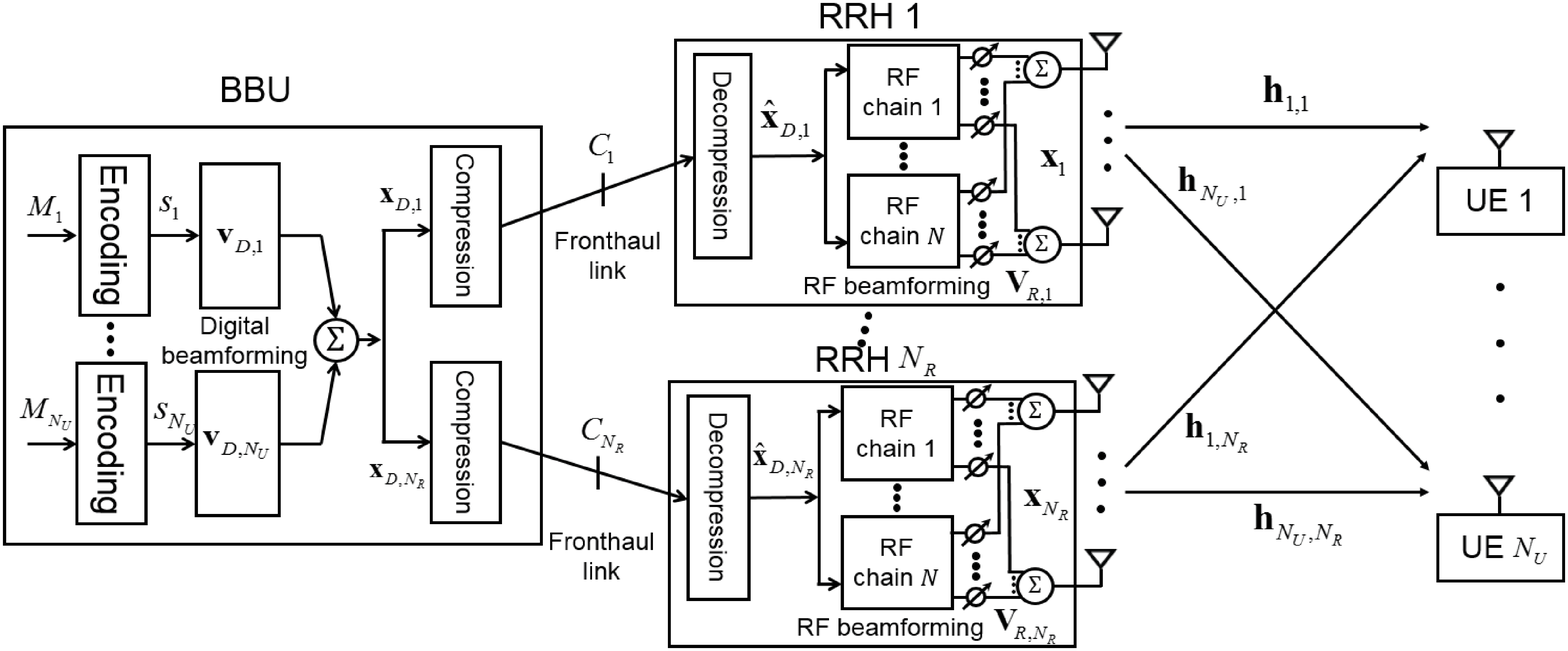}
\end{center}
\vspace{-5mm}
\caption{Illustration of the downlink of a C-RAN with hybrid digital and analog processing.}
\label{figure:Sch_dia}
\end{figure}

In this work, we study the application of hybrid beamforming to the cloud radio access network (C-RAN) architecture (see, e.g., \cite{OSi:16}).
In a C-RAN, the baseband signal processing functionalities of the remote radio heads (RRHs) are migrated to a baseband processing unit (BBU) in the "cloud", which is connected to the RRHs via fronthaul links.
As a result, C-RAN can implement effective interference management techniques owing to centralized cloud processing and can reduce the cost for deploying and operating the RRHs \cite{OSi:16}. However, one of the significant challenges to realize these benefits is the capacity limitation of the fronthaul links that connect the BBU to the RRHs.
To resolve this issue, some advanced fronthaul compression techniques were studied in, e.g., \cite{SHPark:14} and \cite{Zhou:16}.

In this work, we tackle the problem of jointly optimizing digital baseband beamforming and fronthaul compression strategies at the BBU along with RF beamforming at the RRHs with the goal of maximizing the weighted downlink sum-rate.
Fronthaul capacity and per-RRH transmit power constraints are imposed, as well as constant modulus constraints on the RF beamforming matrices, which consist of analog phase shifters.
After formulating the problem, which turns out to be non-convex, we propose an iterative algorithm that is based on successive convex approximation and on the relaxation of the constant modulus constraint. Numerical results are provided to validate the effectiveness of the proposed algorithm.

The paper is organized as follows. In Sec. \ref{Sys}, we present the system model for the downlink of a C-RAN with hybrid digital and analog processing and with finite-capacity fronthaul links. We describe the problem of weighted sum-rate maximization in Sec. \ref{Des_Per}, and we propose an iterative algorithm to tackle the problem in Sec. \ref{sec:optimization}. Numerical results are presented in~Sec.~\ref{Sim_Res}, and the paper is closed with the concluding remarks in  Sec.~\ref{Con}.

\textit{Notations:} Boldface uppercase, boldface lowercase and normal letters indicate matrices, vectors and scalars, respectively.
The circularly symmetric complex Gaussian distribution with mean $\boldsymbol{\mu}$ and covariance matrix $\mathbf{R}$ is denoted by $\mathcal{CN}(\boldsymbol{\mu},\mathbf{R})$.
The set of all $M\times N$ complex matrices is denoted as $\mathbb{C}^{M\times N}$, and $\mathbf{A}(i,j)$ denotes the $(i,j)$th element of a matrix $\mathbf{A}$.
The Hermitian transpose of a matrix $\mathbf{A}$ is denoted by $\mathbf{A}^H$.


\section{System Model} \label{Sys}

As illustrated in Fig. \ref{figure:Sch_dia}, we consider the downlink of a C-RAN where a BBU communicates with $N_U$ single-antenna UEs through $N_R$ RRHs, each equipped with $M$ transmit antennas \cite{OSi:16}.
We assume that the $i$th RRH is connected to the BBU via an error-free fronthaul link of capacity $C_i$ bps/Hz \cite{SHPark:14}, and that each RRH is equipped with $N<M$ RF chains due to cost limitations.
This means that fully digital beamforming across the $M$ transmit antennas of each RRH is not allowed \cite{LLi:14,OEAy:14,AAl:15,XYu:16,FSo:16,CSL:16}.
For convenience, we define the notations $\mathcal{R}\triangleq\{1,\ldots,N_R\}$, $\mathcal{K}\triangleq\{1,\ldots,N_U\}$, $\bar{M}\triangleq N_R M$, and $\bar{N}\triangleq N_R N$.

We assume that the signal processing strategies of the BBU, the RRHs and the UEs are jointly managed by the BBU based on the global channel state information (CSI) reported from the RRHs by means of the fronthaul links. Each RRH can obtain its local CSI via feedback on the uplink in a frequency division duplexing system \cite{Lim} or uplink channel training in a time-division duplexing (TDD) system \cite{Guvensen}. In this work, we focus on the perfect CSI case such that the CSI reported to the BBU is accurate.
We note that this assumption should be revisited in light of the limited number of RF chains, which limit the capabilities of the BBU to estimate the CSI.
Nevertheless, we make here the assumption of perfect CSI, as done in many related references such as \cite{SPark:16} and references therein, and leave the analysis of the effect of imperfect CSI to future~work.

\subsection{Channel Model} \label{sec:channel-model}

For the downlink channel from the RRHs to the UEs, we assume a frequency-flat fading channel model such that the received signal $y_k$ of the $k$th UE is given as
\bea\label{y_k_1}
y_k=\sum_{i\in\mathcal{R}}\mathbf{h}_{k,i}^H \mathbf{x}_i+z_k=\mathbf{h}_k^H \mathbf{x}+z_k,
\eea
where $\mathbf{x}_i\in\mathbb{C}^{M\times1}$ is the transmitted signal of the $i$th RRH, which is subject to the transmit power constraint $E||\mathbf{x}_i||^2\leq P_i$;
$\mathbf{h}_{k,i}\in\mathbb{C}^{M\times1}$ is the channel vector from the $i$th RRH to the $k$th UE;
$\mathbf{x}=[\mathbf{x}_1^H~\dots~\mathbf{x}_{N_R}^H]^H\in\mathbb{C}^{\bar{M}\times1}$ represents the signal vector transmitted by all the RRHs; $\mathbf{h}_k=[\mathbf{h}_{k,i}^H~\dots~\mathbf{h}_{k,N_R}^H]^H\in\mathbb{C}^{\bar{M}\times1}$ indicates the channel vector from all the RRHs to the $k$th UE;
and $z_k\sim\mathcal{CN}(0,1)$ denotes the additive noise at the $k$th UE.



\subsection{Digital Beamforming and Fronthaul Compression} \label{sec:digital-processing}

We denote the message intended for the $k$th UE as $M_k\in\{1,\dots,2^{nR_k}\}$, where $n$ indicates the coding block length and $R_k$ is the rate of the message $M_k$.
The BBU encodes the messages $M_k$ to produce encoded baseband signals $s_k\sim\mathcal{CN}(0,1)$ for $k\in\mathcal{K}$.
Then, in order to handle the inter-UE interference signals, the signals $\{s_k\}_{k\in{\mathcal{K}}}$ are linearly precoded as
\bea
\mathbf{x}_D=\bigg[\mathbf{x}_{D,1}^H~\dots\mathbf{x}_{D,N_R}^H\bigg]^H=\sum_{k\in\mathcal{K}}\mathbf{v}_{D,k}s_k,
\eea
where $\mathbf{v}_{D,k}\in\mathbb{C}^{\bar{N}\times1}$ is the \textit{digital beamforming} vector across all the RRHs for the $k$th UE, and $\mathbf{x}_{D,i}\in\mathbb{C}^{N\times1}$ is the $i$th subvector of $\mathbf{x}_D\in\mathbb{C}^{\bar{N}\times1}$ corresponding to the $i$th RRH.
If we define the shaping matrices $\mathbf{E}_i=[\mathbf{0}_{N\times N(i-1)}^H \, \mathbf{I}_{N} \, \mathbf{0}_{N\times N(N_R-i)}^H]^H$, with $\mathbf{I}_d$ denoting the identity matrix of size $d$, the $i$th subvector $\mathbf{x}_{D,i}$ can be expressed as $\mathbf{x}_{D,i}=\sum_{k\in\mathcal{K}}\mathbf{E}_i^H \mathbf{v}_{D,k}s_k$.


Since the BBU communicates with the $i$th RRH via the fronthaul link of finite capacity $C_i$, the signal $\mathbf{x}_{D,i}$ is quantized and compressed prior to being transferred to the RRH.
Following the references \cite{OSi:16,SHPark:14,Zhou:16}, we model the impact of the compression by writing the quantized signal $\hat{\mathbf{x}}_{D,i}$ as
\bea
\hat{\mathbf{x}}_{D,i}=\mathbf{x}_{D,i}+\mathbf{q}_i,
\eea
where the quantization noise $\mathbf{q}_i\in\mathbb{C}^{N\times1}$ is independent of the signal $\mathbf{x}_{D,i}$ and distributed as $\mathbf{q}_i\sim\mathcal{CN}(\mathbf{0},\mathbf{\Omega}_i)$.
From \cite[Ch.~3]{AEGamal:12}, the quantized signal $\hat{\mathbf{x}}_{D,i}$ can be reliably recovered at the $i$th RRH if the condition
\bea
g_i(\mathbf{V}_D,\mathbf{\Omega}_i)&\triangleq&I(\mathbf{x}_{D,i};\hat{\mathbf{x}}_{D,i})\label{g_i}\\
&=&\log_2\det\bigg(\sum_{k\in\mathcal{K}}\mathbf{E}_i^H\mathbf{v}_{D,k}\mathbf{v}_{D,k}^H\mathbf{E}_i+\mathbf{\Omega}_i\bigg)\nonumber\\
&~&-\log_2\det(\mathbf{\Omega}_i)\leq C_i\nonumber
\eea
is satisfied, where we defined the set of the digital beamforming vectors as $\mathbf{V}_D\triangleq\{\mathbf{v}_{D,k}\}_{k\in\mathcal{K}}$.

\subsection{RF Beamforming}

The quantized signal vector $\hat{\mathbf{x}}_{D,i}$ decompressed at the $i$th RRH is of dimension $N$, which is less than the number $M$ of transmit antennas.
The $i$th RRH applies \textit{analog RF beamforming} to $\hat{\mathbf{x}}_{D,i}$ via a beamforming matrix $\mathbf{V}_{R,i}\in\mathbb{C}^{M\times N}$, so that the transmitted signal $\mathbf{x}_i$ from the $M$ transmit antennas is given as
\bea\label{x_i_1}
\mathbf{x}_i=\mathbf{V}_{R,i}\hat{\mathbf{x}}_{D,i}=\sum_{k\in\mathcal{K}}\mathbf{V}_{R,i}\mathbf{E}_i^H\mathbf{v}_{D,k}s_k+\mathbf{V}_{R,i}\mathbf{q}_i.
\eea
Typically, the RF beamformers are implemented using analog phase shifters (see, e.g., \cite{FSo:16}), and hence the $(a,b)$th element of the RF beamforming matrix $\mathbf{V}_{R,i}$ has the form of $\mathbf{V}_{R,i}(a,b)=e^{j\theta_{i,a,b}}$ for $i\in\mathcal{R}$, $a\in\mathcal{M}\triangleq\{1,\ldots,M\}$, and $b\in\mathcal{N}\triangleq\{1,\ldots,N\}$, where $\theta_{i,a,b}$ indicates the phase shift applied between the signals $\hat{\mathbf{x}}_{D,i}(b)$ and $\mathbf{x}_{i}(a)$.
Therefore, when designing the RF beamforming matrices $\mathbf{V}_{R,i}$, one should satisfy the constant modulus constraints $|\mathbf{V}_{R,i}(a,b)|^2=1$  for all $a\in\mathcal{M}$ and $b\in\mathcal{N}$ (see, e.g., \cite{OEAy:14,XYu:16,FSo:16})).


\section{Problem Description} \label{Des_Per}

In this section, we discuss the problem of jointly designing the  beamforming matrices $\mathbf{V}_R$ and $\mathbf{V}_D$ for the RF and digital parts, respectively, along with the quantization noise covariance matrices $\mathbf{\Omega}$, where we define the notations $\mathbf{V}_R\triangleq\{\mathbf{V}_{R,i}\}_{i\in\mathcal{R}}$ and $\mathbf{\Omega}\triangleq\{\mathbf{\Omega}_i\}_{i\in\mathcal{R}}$.
To measure the achievable rate for each UE $k$, we rewrite the signal $y_k$ in (\ref{y_k_1}) under the transmission model (\ref{x_i_1}) as
\bea\label{y_k_2}
y_k=\sum_{l\in\mathcal{K}} \mathbf{h}_k^H\bar{\mathbf{V}}_R\mathbf{v}_{D,l}s_l+\mathbf{h}_k^H\bar{\mathbf{V}}_R\mathbf{q}+z_k,
\eea
where we defined the effective RF beamforming matrix
$
\bar{\mathbf{V}}_R\triangleq[(\mathbf{V}_{R,1}\mathbf{E}_1^H)^H\dots(\mathbf{V}_{R,N_R}\mathbf{E}_{N_R}^H)^H]^H
$ for all RRHs
and the vector $\mathbf{q}\triangleq [\mathbf{q}_1^H \dots \mathbf{q}_{N_R}^H]^H\in\mathbb{C}^{\bar{N}\times 1}$ of all the quantization noise signals distributed as $\mathbf{q}\sim\mathcal{CN}(\mathbf{0},\bar{\mathbf{\Omega}})$ with $\bar{\mathbf{\Omega}}\triangleq\text{diag}(\mathbf{\Omega}_1,\ldots,\mathbf{\Omega}_{N_R})$.

If we assume that each UE $k$ decodes the message $M_k$ by treating the interference signals as additive noise, the achievable rate $R_k$ for the UE is given as
\bea
R_k&=&f_k(\mathbf{V}_R,\mathbf{V}_D,\mathbf{\Omega})=I(s_k;y_k)\\\label{R_k}
&=&\Phi\bigg(|\mathbf{h}_k^H\bar{\mathbf{V}}_R\mathbf{v}_{D,k}|^2,\zeta_k(\mathbf{V}_R,\mathbf{V}_D,\mathbf{\Omega})\bigg),\nonumber
\eea
where we defined the notations $\zeta_k(\mathbf{V}_R,\mathbf{V}_D,\mathbf{\Omega})\triangleq$
$\sum_{l\in\mathcal{K}\setminus\{k\}}|\mathbf{h}_k^H\bar{\mathbf{V}}_R\mathbf{v}_{D,l}|^2+\mathbf{h}_k^H\bar{\mathbf{V}}_R\bar{\mathbf{\Omega}}_i\bar{\mathbf{V}}_R^H\mathbf{h}_k+1$ and $\Phi(\mathbf{A},\mathbf{B})\triangleq\log_2\det(\mathbf{A}+\mathbf{B})-\log_2\det(\mathbf{B})$.

In this work, we tackle the problem of maximizing the weighted sum-rate $\sum_{k\in\mathcal{K}} w_k R_k$ of the UEs while satisfying the per-RRH transmit power, fronthaul capacity and constant modulus constraints.
The problem is stated as
\begin{subequations}\label{P_1}
\begin{align}
\underset{\mathbf{V}_R,\mathbf{V}_D,\mathbf{\Omega}}{\text{maximize}}&\sum_{k\in\mathcal{K}}w_k f_k(\mathbf{V}_R,\mathbf{V}_D,\mathbf{\Omega})\label{P_1_ob}\\
\text{s.t.~}&g_i(\mathbf{V}_D,\mathbf{\Omega}_i)\leq C_i,~i\in\mathcal{R},\label{P_1_g_i}\\
&p_i(\mathbf{V}_{R,i},\mathbf{V}_D,\mathbf{\Omega}_i)\leq P_i,~i\in\mathcal{R},\label{P_1_p_i}\\
&|\mathbf{V}_{R,i}(a,b)|^2=1,~a\in\mathcal{M},~b\in\mathcal{N},~i\in\mathcal{R}.\label{P_1_modul}
\end{align}
\end{subequations}
The problem (\ref{P_1}) is non-convex due to the objective function (\ref{P_1_ob}) and the constraints (\ref{P_1_g_i}) and (\ref{P_1_modul}).
In the next section, we present an iterative algorithm that obtains an efficient solution.


\section{Proposed Optimization Algorithm} \label{sec:optimization}

In this section, to tackle the problem (\ref{P_1}), we propose an iterative algorithm based on a block coordinate descent approach \cite{DPBer:97} whereby the RF beamforming matrix $\mathbf{V}_R$ and the digital processing strategies $\{\mathbf{V}_D$, $\mathbf{\Omega}\}$ are alternately optimized.
To this end, we describe in Sec. \ref{Opt_Dig} the optimization of the digital part $\mathbf{V}_D$ and $\mathbf{\Omega}$ for fixed RF beamforming $\mathbf{V}_R$, and then present the optimization of the latter in Sec. \ref{sec:opt-RF}.

\subsection{Optimization of Digital Beamforming and Compression} \label{Opt_Dig}

The problem (\ref{P_1}) with respect to the digital beamforming $\mathbf{V}_D$ and fronthaul compression strategies $\mathbf{\Omega}$ for fixed RF beamforming $\mathbf{V}_R=\mathbf{V'}_R$ can be written as
\begin{subequations}\label{P_D}
\begin{align}
\underset{\mathbf{V}_D,\mathbf{\Omega}}{\text{maximize}}&\sum_{k\in\mathcal{K}}w_k f_k(\mathbf{V'}_R,\mathbf{V}_D,\mathbf{\Omega})\label{P_D_ob}\\
\text{s.t.~}&g_i(\mathbf{V}_D,\mathbf{\Omega}_i)\leq C_i,~i\in\mathcal{R},\label{P_D_g_i}\\
&p_i(\mathbf{V'}_{R,i},\mathbf{V}_D,\mathbf{\Omega}_i)\leq P_i,~i\in\mathcal{R},\label{P_D_p_i}
\end{align}
\end{subequations}
where the RF variables $\mathbf{V}_R=\mathbf{V'}_R$ are treated as constants.
The problem (\ref{P_D}) is non-convex due to the objective function (\ref{P_D_ob}) and the constraint (\ref{P_D_g_i}).

To address this, we adapt the algorithm proposed in \cite{YZhou:16}, which is based on successive convex approximation, to the problem at hand.
To this end, we consider a lower bound of the function $f_k(\mathbf{V'}_R,\mathbf{V}_D,\mathbf{\Omega})$ as
\bea
f_k(\mathbf{V'}_R,\mathbf{V}_D,\mathbf{\Omega})\geq\frac{1}{\ln 2}\gamma_k(\mathbf{V'}_R,\mathbf{V}_D,\mathbf{\Omega},u_k,\tilde{w}_k),\label{f_k}
\eea
with arbitrary $\tilde{w}_k\geq0$ and $u_k$, where the function $\gamma_k(\mathbf{V'}_R,\mathbf{V}_D,\mathbf{\Omega},u_k,\tilde{w}_k)$ is defined as
\bea
\gamma_k(\mathbf{V'}_R,\mathbf{V}_D,\mathbf{\Omega},u_k,\tilde{w}_k)~~~~~~~~~~~~~~~~~~~~~&\\
=\ln\tilde{w}_k-\tilde{w}e_k(\mathbf{V'}_R,\mathbf{V}_D,\mathbf{\Omega},u_k)+1,&\nonumber
\eea
with the mean squared error function $e_k(\mathbf{V'}_R,\mathbf{V}_D,\mathbf{\Omega},u_k)$ given as
\bea
e_k(\mathbf{V'}_R,\mathbf{V}_D,\mathbf{\Omega},u_k)&=&|1-u_k^*\mathbf{h}_k^H\bar{\mathbf{V'}}_R\mathbf{v}_{D,k}|^2\\
&+&|u_k|^2\zeta_k(\mathbf{V'}_R,\mathbf{V}_D,\mathbf{\Omega}).\nonumber
\eea
Note that the lower bound in (\ref{f_k}) is satisfied with equality when the variables $u_k$ and $\tilde{w}_k$ are given as
\bea
\!\!\!\!u_k\!\!\!\!&=&\!\!\!\!\!\!\!\bigg(\!\!|\mathbf{h}_k^H\bar{\mathbf{V'}}_R\mathbf{v}_{D,k}|^2\!+\!\zeta_k(\mathbf{V'}_R,\!\mathbf{V}_D,\!\mathbf{\Omega})\!\!\bigg)^{\!\!\!-1}\!\!\!\!\mathbf{h}_k^H\bar{\mathbf{V'}}_R\mathbf{v}_{D,k},\label{u_k}\\
\tilde{w}_k\!\!\!\!&=&\!\!\!\!e_k(\mathbf{V'}_R,\mathbf{V}_D,\mathbf{\Omega},u_k)^{-1}.\label{w_k}
\eea

We also consider an upper bound of the function $g_i(\mathbf{V}_D,\mathbf{\Omega}_i)$ in the constraints (\ref{P_SCA_C1}) as \cite{YZhou:16}
\bea
g_i(\mathbf{V}_D,\mathbf{\Omega}_i)\leq\tilde{g}_i(\mathbf{V}_D,\mathbf{\Omega}_i,\mathbf{\Sigma}_i)\label{g_i_c}
\eea
for arbitrary positive definite matrix $\mathbf{\Sigma}_i$, where we defined
\bea
\tilde{g}_i(\mathbf{V}_D,\mathbf{\Omega}_i,\mathbf{\Sigma}_i)&=&\log_2\det(\mathbf{\Sigma}_i)\\
&+&\!\!\!\!\!\frac{1}{\ln{2}}\text{tr}\bigg(\mathbf{\Sigma}_i^{-1}\bigg(\sum_{k\in\mathcal{K}}\mathbf{E}_i^H\mathbf{v}_{D,k}\mathbf{v}_{D,k}^H\mathbf{E}_i+\mathbf{\Omega}_i\bigg)\bigg)\nonumber\\
&-&\frac{1}{\ln{2}}N-\log_2\det(\mathbf{\Omega}_i).\nonumber
\eea
The matrix $\mathbf{\Sigma}_i$ that makes the inequality (\ref{g_i_c}) tight is given as
\bea
\mathbf{\Sigma}_i=\sum_{k\in\mathcal{K}}\mathbf{E}_i^H\mathbf{v}_{D,k}\mathbf{v}_{D,k}^H\mathbf{E}_i+\mathbf{\Omega}_i.\label{sigma_i}
\eea

Based on the inequalities (\ref{f_k}) and (\ref{g_i_c}), as in \cite{YZhou:16}, we formulate the problem
\begin{subequations}\label{P_D_2}
\begin{align}
\underset{\mathbf{V}_D,\mathbf{\Omega},\mathbf{u},\tilde{\mathbf{w}},\mathbf{\Sigma}}{\text{maximize}}&\sum_{k\in\mathcal{K}}\frac{w_k}{\ln2}\gamma_k(\mathbf{V'}_R,\mathbf{V}_D,\mathbf{\Omega},u_k,\tilde{w}_k)\label{P_D_2_ob}\\
&-\rho\sum_{i\in\mathcal{R}}\|\mathbf{\Sigma}_i-\mathbf{\Phi}_i(\mathbf{V}_D,\mathbf{\Omega})\|^2_F\nonumber\\
\text{s.t.~}&\tilde{g}_i(\mathbf{V}_D,\mathbf{\Omega}_i,\mathbf{\Sigma}_i)\leq C_i,~i\in\mathcal{R},\label{P_SCA_C1}\\
&p_i(\mathbf{V'}_{R,i},\mathbf{V}_D,\mathbf{\Omega}_i)\leq P_i,~i\in\mathcal{R},
\end{align}
\end{subequations}
where we defined the notations $\mathbf{u}\triangleq\{u_k\}_{k\in\mathcal{K}},~\tilde{\mathbf{w}}\triangleq\{\tilde{w}_k\}_{k\in\mathcal{K}}$, $\mathbf{\Sigma}\triangleq\{\mathbf{\Sigma}_i\}_{i\in\mathcal{R}}$, and $\mathbf{\Phi}_i(\mathbf{V}_D,\mathbf{\Omega})\triangleq\sum_{k\in\mathcal{K}}\mathbf{E}_i^H\mathbf{v}_{D,k}\mathbf{v}_{D,k}^H\mathbf{E}_i+\mathbf{\Omega}_i$.
The second summand in (\ref{P_D_2_ob}), with a positive constant $\rho$, can be considered as a regularization term that encourages strong convexity properties \cite{YZhou:16}.

We note that the problem is not equivalent to (\ref{P_D}).
Furthermore, although the problem (\ref{P_D_2}) is still non-convex, it is convex with respect to $(\mathbf{V}_D,\mathbf{\Omega})$ when variables ($\mathbf{u}$, $\tilde{\mathbf{w}}$, $\mathbf{\Sigma}$) are fixed and vice versa.
As proved in \cite{YZhou:16}, solving problem (\ref{P_D_2}) alternately over these two sets of variables yields an algorithm, described in Algorithm 1, that is guaranteed to converge to a stationary point of the problem (\ref{P_D}).

\renewcommand{\arraystretch}{1.}
 \begin{center}
 \begin{tabular}{l}
 \hthickline
    Algorithm 1: Algorithm for updating $\mathbf{V}_D$ and $\mathbf{\Omega}$\\
 \hthickline
    1. Initialize the variables $\mathbf{V}_D^{(1)}$ and $\mathbf{\Omega}^{(1)}$ to arbitrary matrices\\
   ~~~satisfying the constraints (\ref{P_D_g_i})-(\ref{P_D_p_i}) and set $t=1$.\\
    2. Update the variables $u_k^{(t+1)}$ according to (\ref{u_k}) with setting\\
   ~~~$\mathbf{V}_D\leftarrow\mathbf{V}_D^{(t)}$ and $\mathbf{\Omega}\leftarrow\mathbf{\Omega}^{(t)}$ for $k\in\mathcal{K}$.\\
    3. Update the variables $\tilde{w}_k^{(t+1)}$ according to (\ref{w_k}) with setting\\
   ~~~$\mathbf{V}_D\leftarrow\mathbf{V}_D^{(t)}$, $\mathbf{\Omega}\leftarrow\mathbf{\Omega}^{(t)}$ and $u_k\leftarrow u_k^{(t+1)}$ for $k\in\mathcal{K}$.\\
    4. Update the variables $\mathbf{\Sigma}_i^{(t+1)}$ according to (\ref{sigma_i}) with setting\\
   ~~~$\mathbf{V}_D\leftarrow\mathbf{V}_D^{(t)}$ and $\mathbf{\Omega}\leftarrow\mathbf{\Omega}^{(t)}$.\\
    5. Update the variables $\mathbf{V}_D^{(t+1)}$ and $\mathbf{\Omega}^{(t+1)}$ as a solution of\\
    ~~~the convex problem\\
 \end{tabular}
 \end{center}
\vspace{-5mm}
\begin{subequations}\label{P_D_Al}
\begin{align}
\underset{\mathbf{V}_D^{(t+1)},\mathbf{\Omega}^{(t+1)}}{\text{maximize}}&\sum_{k\in\mathcal{K}}\frac{w_k}{\ln 2}\gamma_k(\mathbf{V}'_R,\mathbf{V}_D^{(t+1)},\mathbf{\Omega}^{(t+1)},u_k^{(t+1)},\tilde{w}_k^{(t+1)})\nonumber\\
&-\rho\sum_{i\in\mathcal{R}}\|\mathbf{\Sigma}_i-\mathbf{\Phi}_i(\mathbf{V}_D^{(t+1)},\mathbf{\Omega}^{(t+1)})\|^2_F\\
\text{s.t.~}&\tilde{g}_i(\mathbf{V}_D^{(t+1)},\mathbf{\Omega}_i^{(t+1)},\mathbf{\Sigma}_i^{(t+1)})\leq C_i,~i\in\mathcal{R},\\
&p_i(\mathbf{V}'_{R,i},\mathbf{V}_D^{(t+1)},\mathbf{\Omega}_i^{(t+1)})\leq P_i,~i\in\mathcal{R},
\end{align}
\end{subequations}
 \vspace{-10mm}
\renewcommand{\arraystretch}{1.}
 \begin{center}
 \begin{tabular}{l}
    where the variables $\mathbf{u}^{(t+1)}$, $\tilde{\mathbf{w}}^{(t+1)}$ and $\mathbf{\Sigma}_i^{(t+1)}$, which were~~\\
   updated in Steps 2-4, are excluded from the optimization \\
   space.\\
    6. Stop if a convergence criterion is satisfied. Otherwise,\\
   ~~~set $t\leftarrow t+1$ and go back to Step 2.\\
 \hthickline
 \end{tabular}
 \end{center}

\subsection{Optimization of RF Beamforming} \label{sec:opt-RF}

In this subsection, we discuss the optimization of the RF beamformers $\mathbf{V}_R$ for fixed digital variables $\mathbf{V}_D=\mathbf{V'}_D$ and $\mathbf{\Omega}=\mathbf{\Omega'}$. The problem can be stated as
\begin{subequations}\label{P_RF}
\begin{align}
\underset{\mathbf{V}_R,\mathbf{u},\tilde{\mathbf{w}}}{\text{maximize}}&\sum_{k\in\mathcal{K}}\frac{w_k}{\ln2}\gamma_k(\mathbf{V}_R,\mathbf{V'}_D,\mathbf{\Omega'},u_k,\tilde{w}_k)\label{P_RF_ob}\\
\text{s.t.~}&p_i(\mathbf{V}_{R,i},\mathbf{V'}_D,\mathbf{\Omega'}_i)\leq P_i,~i\in\mathcal{R},\label{P_RF_p_i}\\
&|\mathbf{V}_{R,i}(a,b)|^2=1,~a\in\mathcal{M},~b\in\mathcal{N},~i\in\mathcal{R},\label{P_RF_modul}
\end{align}
\end{subequations}
where we used the lower bound (\ref{f_k}) and have removed the fronthaul capacity constraints (\ref{P_1_g_i}) which do not depend on the RF beamforming variables $\mathbf{V}_R$.

We note that the presence of the constant modulus constraint (\ref{P_RF_modul}) makes it difficult to solve the problem (\ref{P_RF}). To address this issue, as in \cite[Sec. III-A]{CSL:16},
we tackle the problem by relaxing the condition (\ref{P_RF_modul}) to the convex constraints $|\mathbf{V}_{R,i}(a,b)|^2\leq 1$.
Then, we can handle the obtained problem by using a similar approach to Algorithm 1. The detailed algorithm can be found in Algorithm 2.

Since the RF beamforming matrices obtained from Algorithm 2, denoted as $\tilde{\mathbf{V}}_R$, may not satisfy the constraints (\ref{P_RF_modul}), we propose to obtain a feasible RF beamformer $\mathbf{V}_{R,i}$ by projecting $\tilde{\mathbf{V}}_R$ onto the feasible space \cite[Sec. III-A]{CSL:16}. In particular, we find the RF beamformer $\mathbf{V}_{R,i}$ such that the distance $\|\mathbf{V}_{R,i}-\tilde{\mathbf{V}}_{R,i}\|^2_F$ is minimized.
As a result, the beamformer $\mathbf{V}_{R,i}$ is obtained as $\mathbf{V}_{R,i}(a,b)\leftarrow\text{exp}(j\angle\tilde{\mathbf{V}}_{R,i}(a,b))$ for $a\in\mathcal{M}$, $b\in\mathcal{N}$ and $i\in\mathcal{R}$ \cite[Eq. (14)]{CSL:16}.

\renewcommand{\arraystretch}{1.}
\begin{center}
\begin{tabular}{l}
\hthickline
   Algorithm 2: Algorithm for updating $\mathbf{V}_R$\\
\hthickline
   1. Initialize the RF beamforming matrices$\mathbf{V}_R^{(1)}$ satisfying\\
   ~~~the constraints (\ref{P_RF_p_i})-(\ref{P_RF_modul}) and set $t=1$.\\
   2. Update the variables $u_k^{(t+1)}$ according to (\ref{u_k}) with\\
   ~~~setting $\mathbf{V}_R\leftarrow\mathbf{V}_R^{(t)}$ for $k\in\mathcal{K}$.\\
   3. Update the variables $\tilde{w}_k^{(t+1)}$ according to (\ref{w_k}) with\\
   ~~~setting $\mathbf{V}_R\leftarrow\mathbf{V}_R^{(t)}$ and $u_k\leftarrow u_k^{(t+1)}$ for $k\in\mathcal{K}$.\\
   4. Update the RF beamforming matrices $\mathbf{V}_R^{(t+1)}$ as\\
    ~~~a solution to the convex problem\\
\end{tabular}
\end{center}
\vspace{-5mm}
\begin{subequations}
\begin{align}
\underset{\mathbf{V}_R^{(t+1)}}{\text{maximize}}&\sum_{k\in\mathcal{K}}\frac{w_k}{\ln 2}\gamma_k(\mathbf{V}_R^{(t+1)},\mathbf{V}'_D,\mathbf{\Omega}',u_k^{(t+1)},\tilde{w}_k^{(t+1)})\\
\text{s.t.~}&p_i(\mathbf{V}_{R,i}^{(t+1)},\mathbf{V}'_D,\mathbf{\Omega}'_i)\leq P_i,~i\in\mathcal{R},\\
&|\mathbf{V}_{R,i}^{(t+1)}(a,b)|^2\leq 1,~a\in\mathcal{M},~b\in\mathcal{N},~i\in\mathcal{R}.
\end{align}
\end{subequations}
\vspace{-7mm}
\renewcommand{\arraystretch}{1.}
\begin{center}
\begin{tabular}{l}
   6. Stop if a convergence criterion is satisfied.~~~~~~~~~~~~~~~
   \\~~~Otherwise, set $t\leftarrow t+1$ and go back to Step 2.\\
\hthickline
\end{tabular}
\end{center}

In summary, for joint design of the digital beamfomring $\mathbf{V}_D$, fronthaul compression $\mathbf{\Omega}$ and RF beamforming strategies $\mathbf{V}_R$, we propose to run Algorithms 1 and 2 alternately until convergence.
The effectiveness of the proposed algorithm will be confirmed by numerical results in Sec. \ref{Sim_Res}.

$\quad$

\section{Numerical Results} \label{Sim_Res}
In this section, we present numerical results to validate the effectiveness of the proposed joint design of the RF and digital processing strategies.
Throughout the simulations, we consider the cases of $N_R=2$ RRHs, $N_U=8$ UEs and $M=10$ RRH antennas, and evaluate the unweighted sum-rate of the UEs (i.e., $w_k=1$ for all $k\in\mathcal{K}$).

Following \cite{AAdhi:13}, we consider a half wavelength-spaced uniform linear antenna arrays of the RRH antennas such that each channel vector $\mathbf{h}_{k,i}$ is distributed as $\mathbf{h}_{k,i}\sim\mathcal{CN}(\mathbf{0}, \mathbf{R}_{k,i})$.
Here the channel covariance matrix $\mathbf{R}_{k,i}$ is given as
\bea
\mathbf{R}_{k,i}(a,b)=\frac{1}{2\Delta_{k,i}}\int_{\theta_{k,i}-\Delta_{k,i}}^{\theta_{k,i}+\Delta_{k,i}}e^{-j\pi(a-b)sin(\phi)}d\phi,
\eea
where the angle of arrival $\theta_{k,i}$ and the angular spread $\Delta_{k,i}$ are obtained from the distributions $\theta_{k,i}\sim U[-\frac{\pi}{3},\frac{\pi}{3}]$ and $\Delta_{k,i}\sim U[\frac{\pi}{18},\frac{2\pi}{9}]$, respectively.
The notation $U(A,B)$ represents the uniform distribution between $A$ and $B$.

For comparison, we consider two baseline schemes:
\begin{itemize}
\item \textit{Fully digital}: Fully digital beamforming is carried out across the RRH antennas, i.e., $M=N$;
\item \textit{Random RF and optimized digital}: The phases of the RF beamforming matrices are randomly obtained from independent and identically distributed uniform distribution.
\end{itemize}

\begin{figure}
\begin{center}
\includegraphics[width=2.8 in]{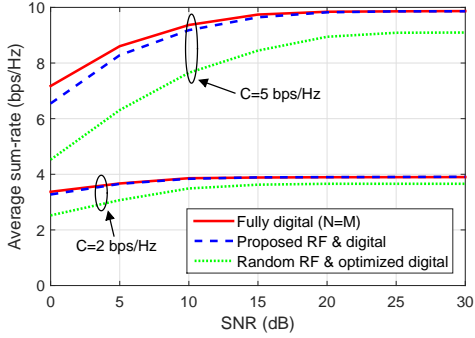}
\end{center}
\vspace{-5mm}
\caption{Average sum-rate performance versus the SNR for the downlink of C-RAN with $N_U=8$, $N_R=2$, $N=2$, $M=10$ and $C\in\{2, 5\}$.}
\label{figure:SNR_perfect}
\end{figure}

Fig. \ref{figure:SNR_perfect} shows the average sum-rate versus the signal-to-noise (SNR) for the downlink of a C-RAN with $N=2$ and $C\in\{2, 5\}$ bps/Hz.
The proposed joint design of the RF and digital processing strategies always outperforms the random RF beamforming scheme, particularly at lower SNR, where the downlink channel becomes the bottleneck of the system.
In a similar way, the optimization of the RF beamforming has more impact when the fronthaul capacity is larger.
Also, as the SNR increases, and hence as the fronthaul capacity limitations become the performance bottleneck, the proposed joint design approaches the sum-rate of the fully digital scheme in spite of limited number of RF chains.

\begin{figure}
\begin{center}
\includegraphics[width=2.8 in]{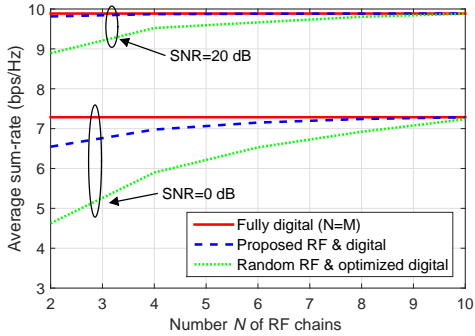}
\end{center}
\vspace{-5mm}
\caption{Average sum-rate performance versus the number $N$ of RF chains for the downlink of C-RAN with $N_U=8$, $N_R=2$, $M=10$, $C=5$ and SNR=0 or 20 dB.}
\label{figure:SR_asN}
\end{figure}

In Fig. \ref{figure:SR_asN}, we plot the average sum-rate versus the number $N$ of RF chains for the downlink of a C-RAN with $C=5$ and $\text{SNR}=0$ or 20 dB. We can see that the proposed joint design shows sum-rate performance that increases more rapidly with $N$ as compared to that of the random RF beamforming scheme.
Also, when $N$ is sufficiently large, both the proposed scheme and random RF beamforming achieve sum-rate performance very close to that of the fully digital beamforming schemes.
Similar to Fig. \ref{figure:SNR_perfect}, the impact of RF beamforming is more pronounced when the SNR is smaller for fixed fronthaul capacity.


\section{Concluding Remarks} \label{Con}
We have studied the joint design of RF and digital signal processing strategies for the downlink of a C-RAN with large-scale antenna arrays. Specifically, we tackled the problem of jointly optimizing the digital beamforming, fronthaul compression and RF beamforming strategies with the goal of maximizing the weighted sum-rate of the UEs, while satisfying the per-RRH power, fronthaul capacity and constant modulus constraints. We have proposed an iterative algorithm that achieves an efficient solution, and we have provided numerical results that validate the advantages of the proposed algorithm.
As an important future work, we mention the development of hybrid processing for a C-RAN with imperfect CSI. Specifically, in a TDD system, the design of RF beamforming should also take into account the goal of CSI acquisition in the uplink channel training (see, e.g., \cite{Guvensen}).



\bibliographystyle{ieeetr}
\bibliography{AZREF}

\end{document}